\documentclass[10pt]{article}

\usepackage{amsmath}
\usepackage{amssymb}

\usepackage{graphicx}

\usepackage{cite}

\usepackage{color}

\usepackage[labelfont=bf,labelsep=period,justification=raggedright]{caption}

\makeatletter
\renewcommand{\@biblabel}[1]{\quad#1.}
\makeatother

\date{}

\begin{document}

\begin{flushleft}
{\Large
\textbf{Formation of Cystine Slipknots in Dimeric Proteins}
}
\\
Mateusz Sikora$^{1}$,
Marek Cieplak$^{1}$,
\\
\bf{1} Institute of Physics, Polish Academy of Sciences, Al. Lotnik\'ow 32/46, 02-668 Warsaw, Poland
\\
$\ast$ E-mail: Corresponding mc@ifpan.edu.pl
\end{flushleft}

\section*{Abstract}
We consider mechanical stability of dimeric and monomeric proteins with
the cystine knot motif. A structure based dynamical model is used to demonstrate
that all dimeric and some monomeric proteins of this kind should have considerable
resistance to stretching that is significantly larger than that of titin.
The mechanisms of the large mechanostability are elucidated.
In most cases, it originates from the induced formation of one or two cystine slipknots.
Since there are four termini in a dimer, there are several ways
of selecting two of them to pull by. We show that in the cystine knot
systems, there is strong anisotropy in mechanostability and force patterns
related to the selection.
We show that the thermodynamic stability of the dimers is enhanced compared to the
constituting monomers whereas machanostability is either lower or higher.


\section*{Introduction}
The cystine knot motif is an interlaced structural arrangement involving three
cystins, i.e. three pairs of cysteines connected by disulfide bonds.
Two of these cystins effectively transform two short segments of the
backbone into a closed ring. The third one connects two different parts of the
backbone through the ring \cite{Sun,Iyer}. This tight structure provides remarkable
thermodynamic stability. It has been first observed in a nerve growth factor (NGF)
\cite{Wlodawer} and then identified in other growth factors \cite{Bradshaw}.
It has also been found in small cysteine-rich toxins \cite{Iyer}, where it was found
to stabilise the structure of small cyclic peptides to a greater extent than the cyclisation of the backbone~\cite{craik1}.
These toxins remain stable and active at temperatures  nearing the boiling point
or at large concentrations of chemical denaturants and enzymes.

The cystine knot motif is highly conserved and is a part of
many growth factors \cite{skeleton1,skeleton2}.
The growth factors are involved in the
development, tissue differentiation and healing processes in all phyla.
In mammals, for instance, there are over 30 different
kinds of TGF-$\beta$ proteins which control these processes.
In particular these proteins can be potential drug targets
in cancer therapy \cite{nickel,napoleon}.

Growth factors are usually flat and extended. They lack a hydrophobic core
and expose the hydrophobic residues to the solvent.
The growth factors typically form dimers \cite{Forstner}. Upon dimerisation,
the exposed hydrophobic residues become burried and generate
attractive contact interactions which bind the monomers in
conjunction with up to two intermeric disulfide bonds.
The dimeric structures are thus rigid and stable in the solvent.
Additionally, residues involved in the cystine knot
form an evolutionaly conserved framework of amino acids responsible
for interactions with receptors of the growth factors \cite{Forstner}.
For instance, in the transforming growth factors-$\beta$ (TGF-$\beta$)
family the receptor binding sites reside near the cystine knot motif
whereas the remaining residues are quite variable and lead to the phenomenon
known as receptor promiscuity~\cite{nickel,Isaacs}.
This phenomenon results in the ability to
bind to distinct receptors, while still maintaining 
the capacity to dimerize the receptors and initiate the
TGF-$\beta$ signalling pathway \cite{Alberts}.

It should be noted that growth factors are also involved in
mechanical processes. It has been shown \cite{nickel,tenney}, that
the growth factors are secreted by a cell in the form of pre-proteins with
a hydrophobically attached long peptide which intertwines with the
growth factor. The peptide enhances solubility. In the process of
maturation, the peptide appears to detach by mechanical forces
exerted by the extracellular matrix. The released growth factor
interacts with corresponding receptors and via proteins of the SMAD-family
releases transcription factors, influencing activity of the cell.

The mechanical stability of proteins with cystine knots has
not been assessed experimentally yet.
However, theoretical studies of proteins
with cystine knot, based on coarse grained \cite{PLOS} and
all-atom models \cite{Peplowski} indicate that it may be
significantly higher than that of titin or ubiquitin. Specifically,
the characteristic force, $F_{max}$, needed to unravel the tertiary structure
may be in the range of even 1 nN, i.e. about five times as big as for titin \cite{titin},
and yet smaller than a force needed to break a covalent bond \cite{Grandbois,Schmidt}.
For the sake of comparison, $F_{max}$ of between 800 and 900 pN has been reported for the
protein molecule in the spider capture-silk thread \cite{Hansma}.

The parameter $F_{max}$ is determined by stretching a protein at a constant speed and observing
the largest force peak when plotting the force, $F$, against the pulling
spring displacement, $d$.
For most proteins, the force peaks are due to shear
between two or more $\beta$-strands \cite{Crampton,COSTB,NAR}. On the other hand,
for the proteins with cystine knots, $F_{max}$ arises due to formation of a cystine slipknot
that takes place when the ring-piercing disulfide bond drags a segment of the backbone
through the ring.
It should be noted that the cystine slipknot is distinct
from the protein slipknot such as studied in protein AFV3-109 \cite{Hongbin}.
The cystine slipknot is formed dynamically whereas the slipknot in AFV3-109
is present in the native state and pulling gets it untied in multiple pathways.
We have found that the top 13 strongest proteins,
among the 17 134 simulated \cite{PLOS,NAR},
are endowed with the cystine slipknot mechanism. There also a hundred of
such proteins with a smaller mechanostability.
The protocol used in the theoretical
studies involved taking into consideration the first structure which is
associated with a given Protein Data Bank (PDB) code \cite{Berman2000} -- either the
first chain or the first NMR-based model. It has turned out, however, that many
of the cystine slipknot proteins are dimers and that this feature has much bigger
dynamical consequences than in the case of non-cystinic complexes. These
circumstances call for reinvestigation of the behavior of proteins with the
cystine knot during stretching. In this paper, we show that such dimeric proteins
are indeed remarkably stable mechanically, various variants of the slipknot
mechanisms are operational, and that the response to stretching, even the very existence
of force peaks, strongly depends on the choice of the two out of four terminal
amino acids to pull by (non-terminal points of force attachment
add to the variety of choices).
Such a  highly anisotropic dependence on the selection of the termini has been
predicted for the titin Z1Z2-telethonin complex \cite{Schulten,Bertz} and, more
recently \cite{domains}, for the 3$D$
domain-swapped cystatin \cite{jaskolski} -- the dimeric protein without any cystine knots.
In both of these cases, however, the mechanical clamps involved are common as they are due to shear.

\subsection*{Geometry of the systems studied}

Here, we focus on four main families of proteins that contain growth factor
cystine knots (GFCKs) -- one of the three known kinds of cystine knots
(the other being inhibitor cystine knot and cyclic cystine knots)
\cite{Sun,Iyer,gracy}. These are
TGF-$\beta$, NGF, glycoprotein hormones (GPH), and
platelet-derived growth factor (PDGF). The latter family has a branch of vascular
endothelial growth factors (VEGF; its human homologues are known as PlGF -- placenta
growth factors).
Table I lists the specific structures we investigate here, together
with their family affiliations and the values of $F_{max}$ for various ways of
pulling. $F_{max}$ is in units of $\epsilon$/{\AA}, which should be of order
110 pN, where $\epsilon$ is the
depth of the potential well associated with each native contact.
Throughout the paper, we use the theoretical units to allow for an easier
comparison with our surveys.
Table I contains 8 proteins belonging to the TGF family and 6 -- to
the VEGF family. These proteins coincide with the 13 top strength structures
found in the mechanostability survey of monomers \cite{PLOS}.
In addition, we consider one protein, 1TFG, which should have been
at the top part of the list as well if not for its mistaken removal from
consideration by structure filtering algorithms.
(We have excluded from considerations the PDGF BB protein with the structure
code 1PDG, as this structure is incomplete). 
Table I lists
two structures associated with the PDB code 1M4U: chain L and chain A
and hence the corresponding subscripts. 1M4U$_{\rm A}$ is known as noggin
whereas 1M4U$_{\rm L}$ studied in ref. \cite{PLOS} is a ligand of noggin and
is known as BMP-7 protein (bone morphogenetic protein).
1M4U$_{\rm L}$ belongs to the TGF family and its
properties are similar to those of 1BMP. Noggin, on the other hand,
is a signalling protein involved in many developmental processes such as
neural induction and bone development \cite{noggin}. It acts through
inactivation of the ligands belonging to the TGF-$\beta$ family.
Each of the chains L and A separately forms dimers and each monomer contains a cystine ring.
The ring in 1M4U$_{\rm A}$ is wider: it contains 2 more residues than in 1M4U$_{\rm L}$.
If 1M4U$_{\rm A}$ and 1M4U$_{\rm L}$ bind then a tetramer forms.

\begin{figure*}[htb]
\includegraphics[width=0.65\textwidth]{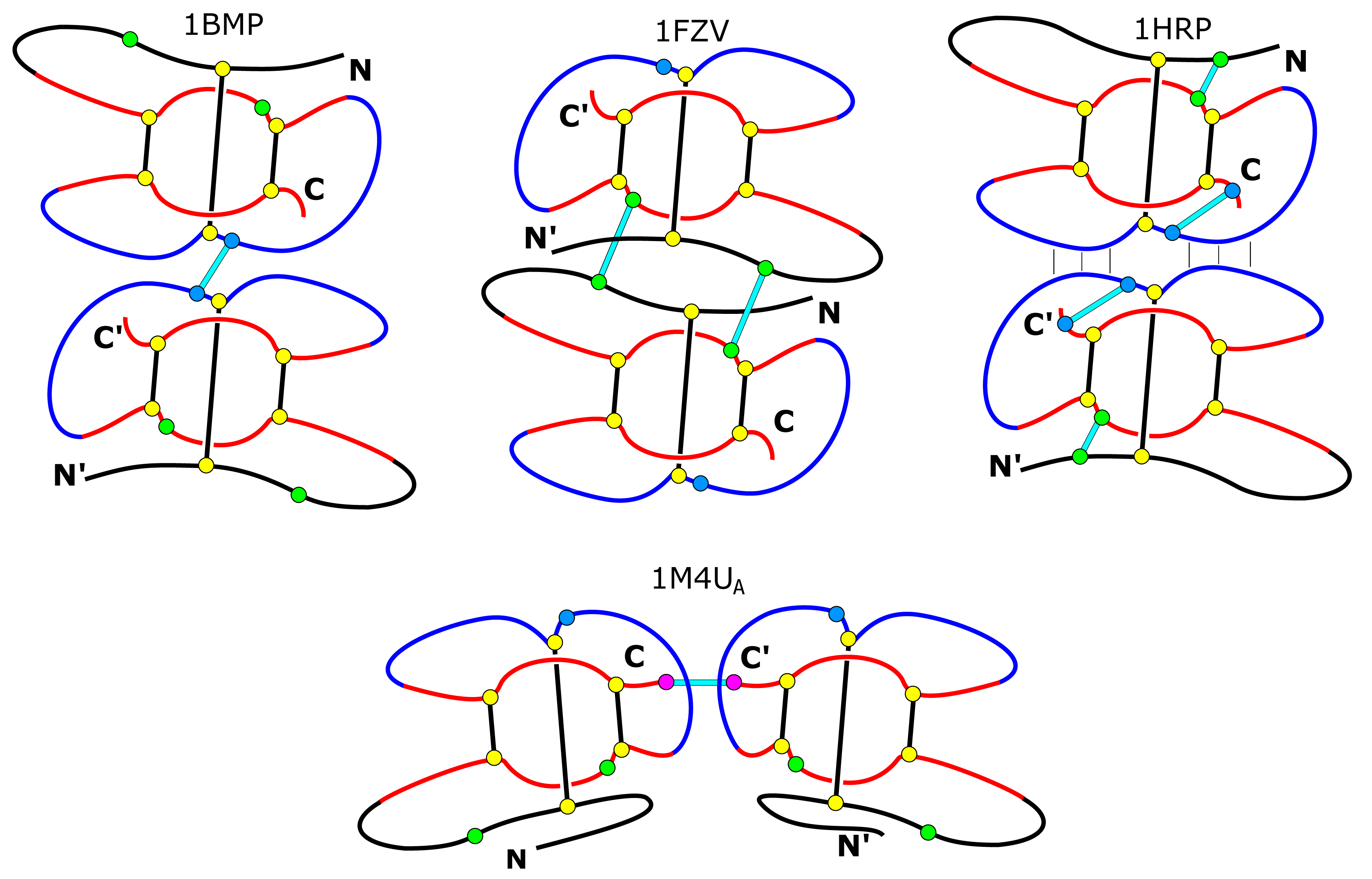}
\caption{{\small Schematic representations of four types of dimer architectures
as exemplified by structures 1BMP and 1FZV, representing the TGF and
VEGF families respectively and structures 1M4U$_{\rm A}$ and 1HRP as indicated.
The cysteines involved in the cystine knot motif are shown as yellow circles.
Other relevant sites are shown as circles either in blue, green, or magenta.
In at least one of the four families shown here, these sites are occupied by
cysteines -- this happens if the circles are connected by lines, i.e. disulfide bonds.
For instance, in 1BMP the green circles  do not show cysteines
but the sites that would house cysteines in the structure corresponding to 1FZV.
The symbols N,C and N', C' in the drawings do not indicate locations
of the terminal amino acids since the corresponding backbones do not end
at these places. Rather, they indicate amino acids which are sequentialy
closest to the indicated termini.
The intra-monomer disulfide bridges are represented by thick black lines, whereas
the inter-monomer bridges are shown as lines in cyan. The monomers in
1BMP and 1M4U$_{\rm A}$ are connected through one cystine but in two different
ways. In 1M4U$_{\rm A}$ the cystine effectively links the rings as it provides
connection of Cys230 on the ring through the nearby C termini Cys232 to Cys230' on the
other ring.
In 1BMP it links
amino acids just next to the ring-piercing cysteins.
In 1FZV there are two binding cystines. Each of them links a ring in one monomer
with an N-proximal segment in the other monomer.
For 1BMP and 1FZV the rings comprise 8 amino acids.
In the case of 1M4U$_{\rm A}$ -- 10 amino acids.
In the case of 1HRP, The vertical lines between two monomers indicate hydrophobic contacts
and hydrogen bonds.}
}\label{dimer_types}
\end{figure*}

Figure \ref{dimer_types} shows schematic connectivities of proteins belonging
to the TGF and VEGF families, as represented by 1BMP and 1FZV structures respectively.
It also shows the scheme of connectivities in 1M4U$_{\rm A}$. In the latter case,
the two monomers are linked through the cysteinic termini C and C'
which are one residue away from a cysteine on the ring.
In the case of 1BMP, the chain-connecting cystine (with Cys103) is just next to
the cysteine (Cys104) that forms a ring-piercing disulfide bond with a cysteine
(Cys38) which is sequentially near the N-terminus (the known structure extends
between sites 36 and 139). The 1FZV dimer has two cystins that bridge the monomers.
Each of them links a ring in one monomer
with an N-proximal segment belonging to the other monomer.

\begin{figure}[htb]
\includegraphics[width=0.7\textwidth]{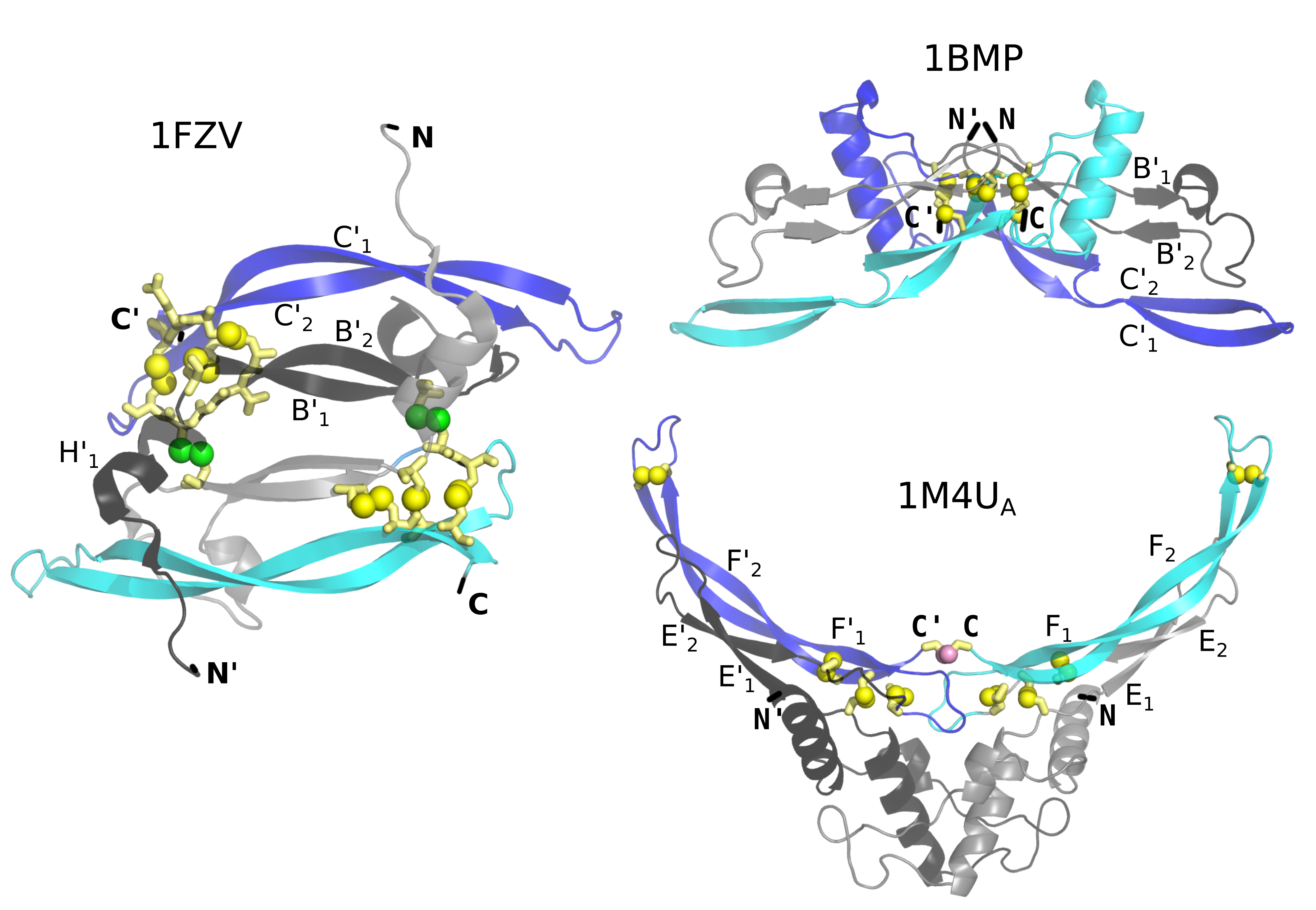}
\caption{{\small The molecular representation of the native dimeric structures
1FZV, BMP, and 1M4U$_{A}$ as shown by the labels. The termini are indicated.
The unprimed symbols refer to one monomer and the primed symbols to the
other. The terminal amino acids are indicated in black. The yellow
spheres correspond to the atoms of sulfur belonging to the cystine rings.
In the panel corresponding to 1FZV, the green spheres correspond to the
cysteines that link the two monomers. In the panel corresponding to 1BMP,
the atoms of sulfur in the cystines that link the two monomers are hidden
behind the yellow spheres. In the panel corresponding to 1M4U$_{A}$,
the inter-monomeric disulfide bond is indicated by two spheres in pink;
one of them is in front of the other.
}}\label{structure_razem}
\end{figure}

More realistic representations of these three native dimeric structures
are shown in figure \ref{structure_razem}.
The overal elongated geometry of a GFCK monomer is well conserved in the
examples shown and is perhaps best seen in top left panel corresponding to 1FZV.
The structure is dominated by a long, 2-stranded $\beta$-sheet
(strands $C_{1}'$ and $C_{2}'$)  that span the whole length of the protein.
It is accompanied by a shorter 2-stranded sheet ($B_{1}'$ and $B_{2}'$).
In the case of 1BMP, both sheets are partitioned into two segments as
seen in the top right panel. The cystine knot is located
close to one of the ends of the monomer and seems to act as as separator
between the sheets. While the geometry of a monomer changes a little between
the specific proteins, we observe that the dimeric proteins employ three
schemes to connect. In the case of the VEGF proteins, the monomers are arranged
antiparallelly, as shown for 1FZV in the top middle panel of figure \ref{dimer_types},
which allows for formation of two connecting disulfide bonds.
In the case of the TGF proteins, the monomers are connected at an angle
and then only by one disulfide bond as illustrated for  1BMP in the top left panel.
Finally, the monomers may be just touching in a small region where
they form a disulfide bond that connects the C and C' termini. This situation
happens in 1M4U$_{\rm A}$, as illustrated in the bottom panel of figure \ref{structure_razem}.

The inter-meric connectivities discussed so far are
provided primarily by disulfide bonds but additional linkages can
arise due to hydrophobicity and hydrogen bonds forming between
the monomers. An alternative
strategy to bind is not to  involve the cystines.
This situation happens for 1HRP and the corresponding connectivities are
indicated in the top right panel of figure \ref{dimer_types}.
A similar pattern is also valid for 1BET.

For completness, we have also studied five monomeric proteins
with the cystine knots. They are listed as the bottom
entries in Table I. One of them, denoted as lefty,
is a member of the  nodal family of signaling factors. It is
expressed during embryo development and is responsible for
formation of the left-right axis \cite{schier}.
The monomeric character of lefty is not certain but it is likely
to be valid \cite{Sakuma}.
The remaining entries are exceptionally stable
short peptides known as knottins, all sharing a
cystine knot motif, but interlaced differently than in the
families of growth factors \cite{gracy}. This motif still contains
a cystine that pierces a ring.
A knottin is formed by the cystine knot that is stripped off any
surrounding secondary structures.
Here, we do not study a subgroup of knottins in which
the backbone is circular which makes the proteins
ultra stable.

\subsection*{The modeling procedure}

The modeling is done
within the coarse grained dynamical model used in ref. \cite{PLOS}, described in more details
in refs. \cite{thermtit,JPCM,models}.
The starting point is to form a polymeric chain of beads that are
tethered together by a harmonic potential. Each bead represents
an amino acid. The disulfide bonds are covalent and are also
represented by the harmonic potentials. We account for the local
backbone stiffness by introducing 4-body terms which favor the
native sense of the local chirality, which is nearly equivalent to
favoring native values of the dihedral angles. The remaining
interactions are defined in terms of contacts: native and non-native.
The native contacts are determined based on the overlap of the
van der Waals spheres assigned to heavy atoms and the $i,i+2$ contacts
are discarded as they are usually weak.
Our structure-based modeling relies on assigning pair-wise
binding potentials to two amino acids that are linked by a native
contact and assigning repulsive potentials to all other pairs
of amino acids, i.e. to the non-native contacts.
We consider a soft repulsive potential which
acts when the distance, $r_{ij}$ between the C$^{\alpha}$
atoms is less than 4 {\AA}. The condition on the binding potentials
is that their minima should correspond to the experimentally
determined native distance between the C$^{\alpha}$ atoms in the
contact-making amino acids. There are countless ways in which
such potentials can be constructed. However, only some of them
lead to unobstructed folding to the native state and to consistency
with the experiments on stretching. We have considered 62
models and tested them against the experimental values of
$F_{max}$ obtained for 38 systems \cite{models}. We have identified
four models which are optimal. Here, we use the simplest of the
four in which the binding potential has the Lennard-Jones form
\begin{equation}
V^{NAT}_{ij}=4\epsilon[(\frac{\sigma_{ij}}{r_{ij}})^{12}-(\frac{\sigma_{ij}}{r_{ij}})^{6}]
\end{equation}
in which the energy parameter $\epsilon$ does not depend on the
identity of the amino acids whereas
the length parameters, $\sigma_{ij}$,
are selected so that the minima of the potentials
agree with the native distances.
The callibration of $\epsilon$ is obtained by finding the best
fit to the experimental data on $F_{max}$ and is given by \cite{PLOS}
$\sim$110 pN$\;${\AA} that has been mentioned in the previous section.
The predictions of this model have been positively verified by stretching
experiments on two scaffoldin proteins \cite{Valbuena}.

The molecular dynamics simulations of stretching are done at temperature $T=0.35 \epsilon/k_B$,
which should correspond to a vicinity of the room temperature
($k_B$ is the Boltzmann constant) and is in the region of good folding.
Stretching is accomplished by attaching
a spring at each of the two selected termini considered. The other end
of one spring is anchored and that of another is made to move with
a constant speed.
The pulling speed is of order 0.005 {\AA}/ns. This speed is within the range
of some stretching experiments \cite{PLOS}, but is some two orders
of magnitude faster than typical stretching exepriments at slower speeds.
Nevertheless it is some five orders of magnitude slower than typical all-atom simulations.
In our regime of speeds, $F_{max}$ depends on the speed merely
logarithmically, so estimates at 0.005 {\AA}/ns are meaningful.
The extrapolation to experimentally accessible pulling speeds yields $F_{max}$
by about 10\% smaller. For instance, we estimate that for 2GH0 the force
of 12.0 $\epsilon$/{\AA} ($\sim$ 1320 pN) gets reduced to 
to 10.45 $\epsilon$/{\AA} ($\sim$ 1150 pN) when lowering the speed from
the theoretical $5\times10^5$ nm/s to the expected experimental speed of 500 nm/s. 
The spring constant is taken to be 0.12 $\epsilon/${\AA}${-2}$ which is of the order
of typical AFM lever elastic constants. We have found \cite{thermtit} that
the choice of the spring constant influences the way the $F-d$ pattern is spread
but it has only a minor effect on the value of $F_{max}$. When studying
extraction of bacteriorhodopsin from a membrane \cite{Janovjak}, an agreement
with the look of the $F-d$ pattern was obtained by reducing the theoretical
spring constant by a factor of 1.35. For each case studied, we have considered
up to ten trajectories and most typical behavior was chosen for a
display in the figures. The value of $F_{max}$ is averaged.

In addition to the mechanical stability, we also assess thermal stability.
The thermal stability of a protein can be characterized by the folding
temperature, $T_f$. It can be defined computationally \cite{Onuchic} in a long
equilibrium run as a temperature
at which the probability, $P_0$, of staying in the native state crosses $\frac{1}{2}$.
One may argue that the system can be considered as staying in the native state
when all of its native contacts are present. However, it is a matter of choice
to declare at which distance between a pair of the C$^{\alpha}$ atoms a contact
is still operational. Our criterion is to take the inflection point in the
contact potential as providing a working threshold.

\section*{Results/Discussion}

\subsection*{Assessment of mechanostability}

For each of the proteins listed in Table I, the termini extend out to
the solvent and could thus be grasped in experiments involving
single molecule manipulation \cite{Nagy,encyclopedia} easily.
If N and C refer to the termini of the first dimer partner, and N' and C'
to the second then there are four choices to select
pairs of termini in which one terminus is anchored and another is pulled.
These choices are indicated
by the symbols N-C', C-C', N-N', and N-C (due to the symmetric arrangement of
cystine rings in both monomers, N'-C' yields same results as N-C, etc.).
In this notation, N-C' means anchoring the N terminus of
the first monomer and pulling by the C-terminus of the second monomer.
Anchoring at C' and pulling at N yields the same result.
The dependence of $F_{max}$ on the choice of attachment points has been
discussed for monomeric proteins, for instance, in refs. \cite{B2,CieMarszalek}.

\begin{figure}[htb]
\includegraphics[width=0.5\textwidth]{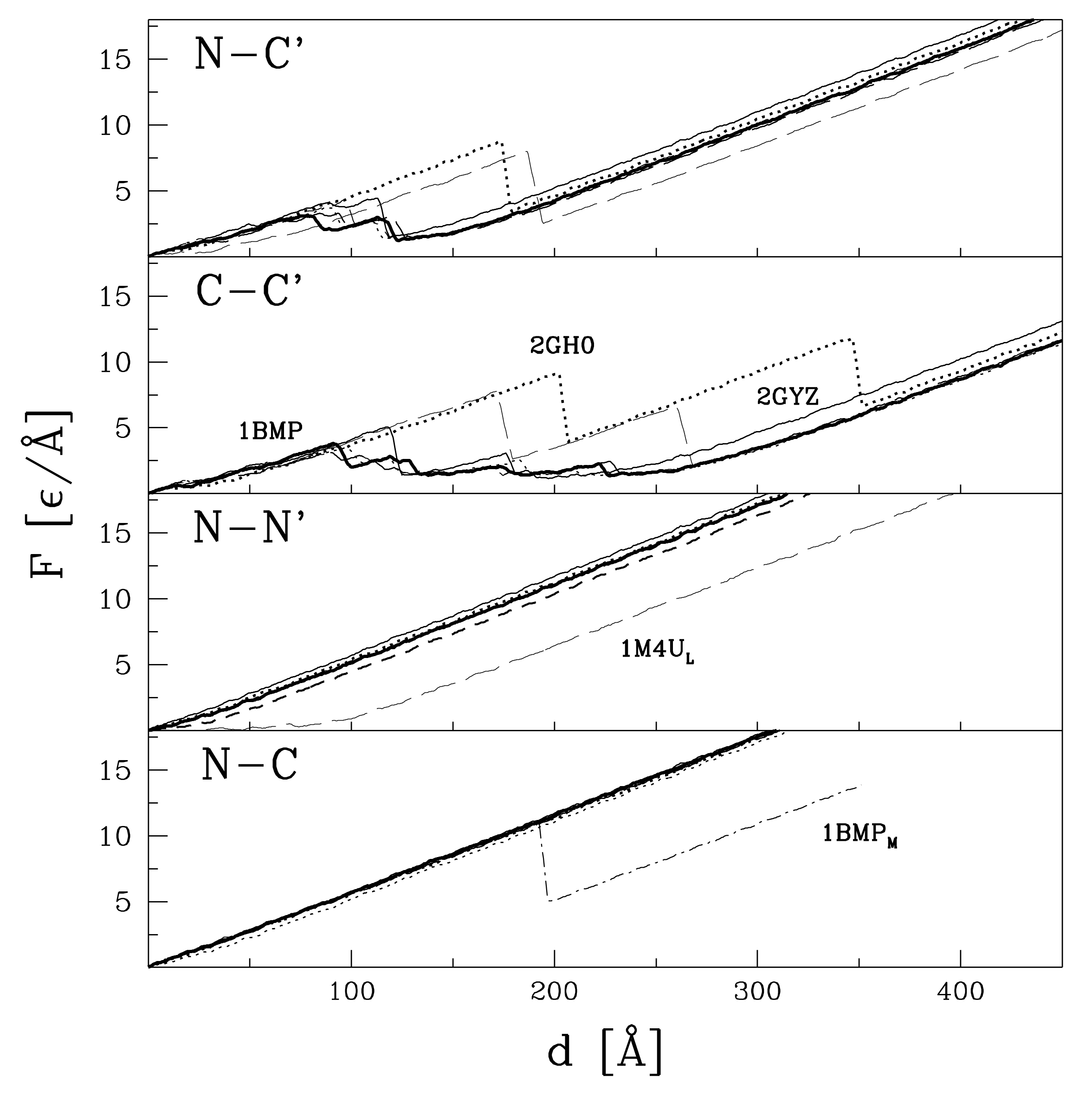}
\caption{ {\small The $F-d$ curves for the proteins of the TGF family that are listed in Table I.
The ways of pulling are indicated in the upper left corner of each panel. The line with
the symbol 1BMP$_{\rm M}$ in the lowest panel indicates the result for a single monomer,
if extracted from the dimer. For other curves, the line type for a given protein is the same
in each panel.}
} \label{typ_1bmp}
\end{figure}

\begin{figure}[htb]
\includegraphics[width=0.5\textwidth]{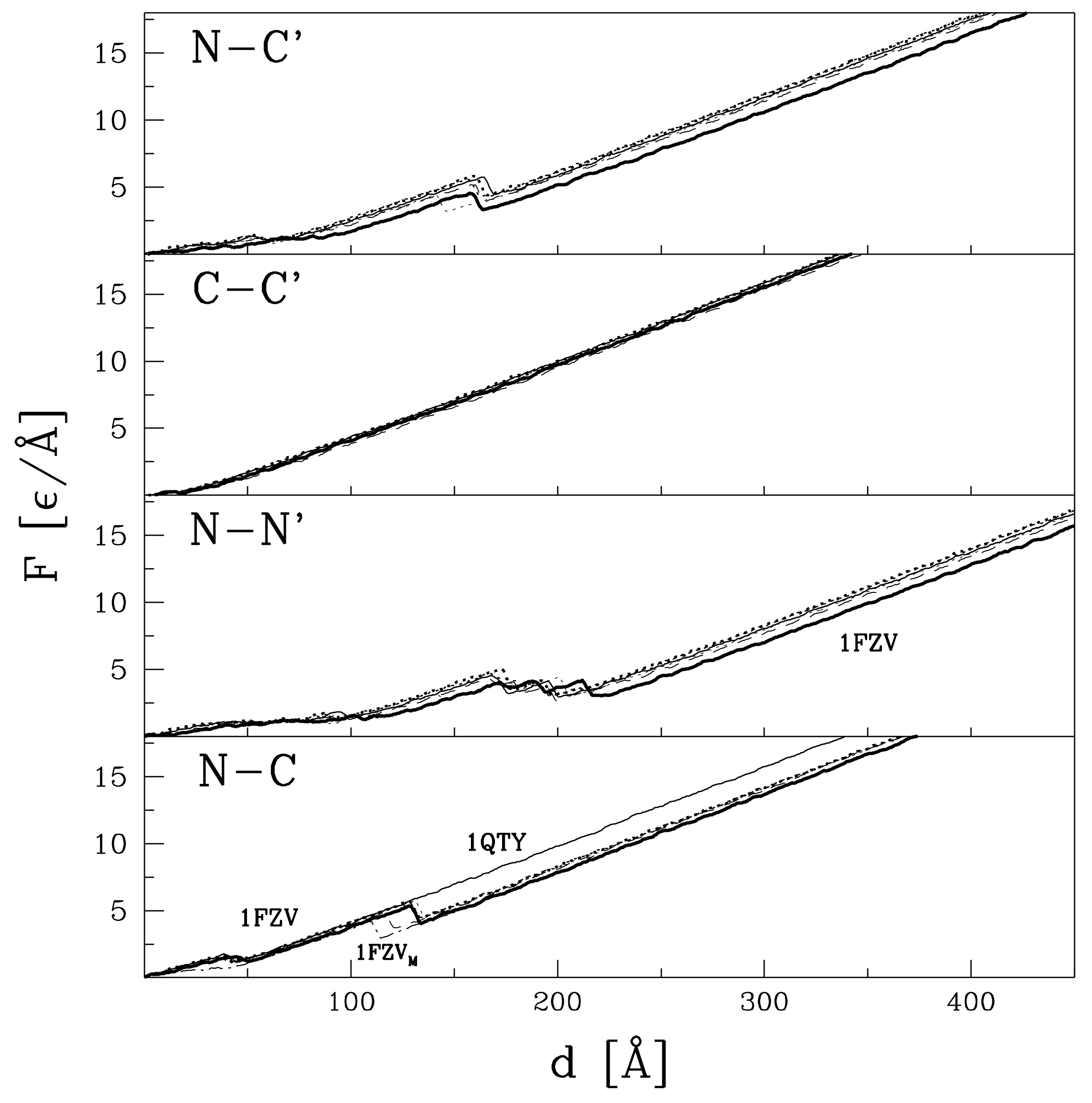}
\caption{{\small Similar to figure \ref{typ_1bmp} but for the VEGF proteins
listed in Table I.}
}\label{typ_1fzv}
\end{figure}

Figures \ref{typ_1bmp} and \ref{typ_1fzv} show the $F-d$ curves for the TGF
and VEGF proteins, that are listed in Table I, respectively. We observe that
for each way of pulling, the curves 
within the TGF family are similar to each other
and so are the curves within the VEGF family.
In particular, there are no force peaks in the TGF proteins
when pulled within the N-N' and N-C schemes and no force peaks in the VEGF
proteins when pulled in the C-C' way. Whenever force peaks do arise,
their heights, shown in Table I, are similar within families and
across families. They are about
5 $\epsilon$/{\AA}, i.e. of order 550 pN. The exception is 2GH0 for which
$F_{max}$ is twice as big.
It should be noted that the $F-d$ curves appear to have essentially no
curvature for large values of $d$, past the last force peak, whereas
a finite curvature is predicted by the worm-like-chain model \cite{worm}.
The reason is that the theoretical model applies to entropic chains
in which potential energy contributions are negligible. This is not the
situation encountered at such high tensions as considered here though some
curvature might become perceptible when going to considerably larger extensions.

Insights into the  nature of the pulling process can be obtained by monitoring
transformations in the conformations. This is illustrated for 1BMP in
Figures \ref{ineffective} and \ref{snap} as well in a movie available
in the Supplementary Information. Figure \ref{ineffective} addresses
situations in which isolated force peaks do not arise (N-C and N-N' pullings)
and figure \ref{snap} is for the C-C' pulling when they do.
Figure \ref{snap} shows six subsequent stages of the process.

\begin{figure}[htb]
\includegraphics[width=0.5\textwidth]{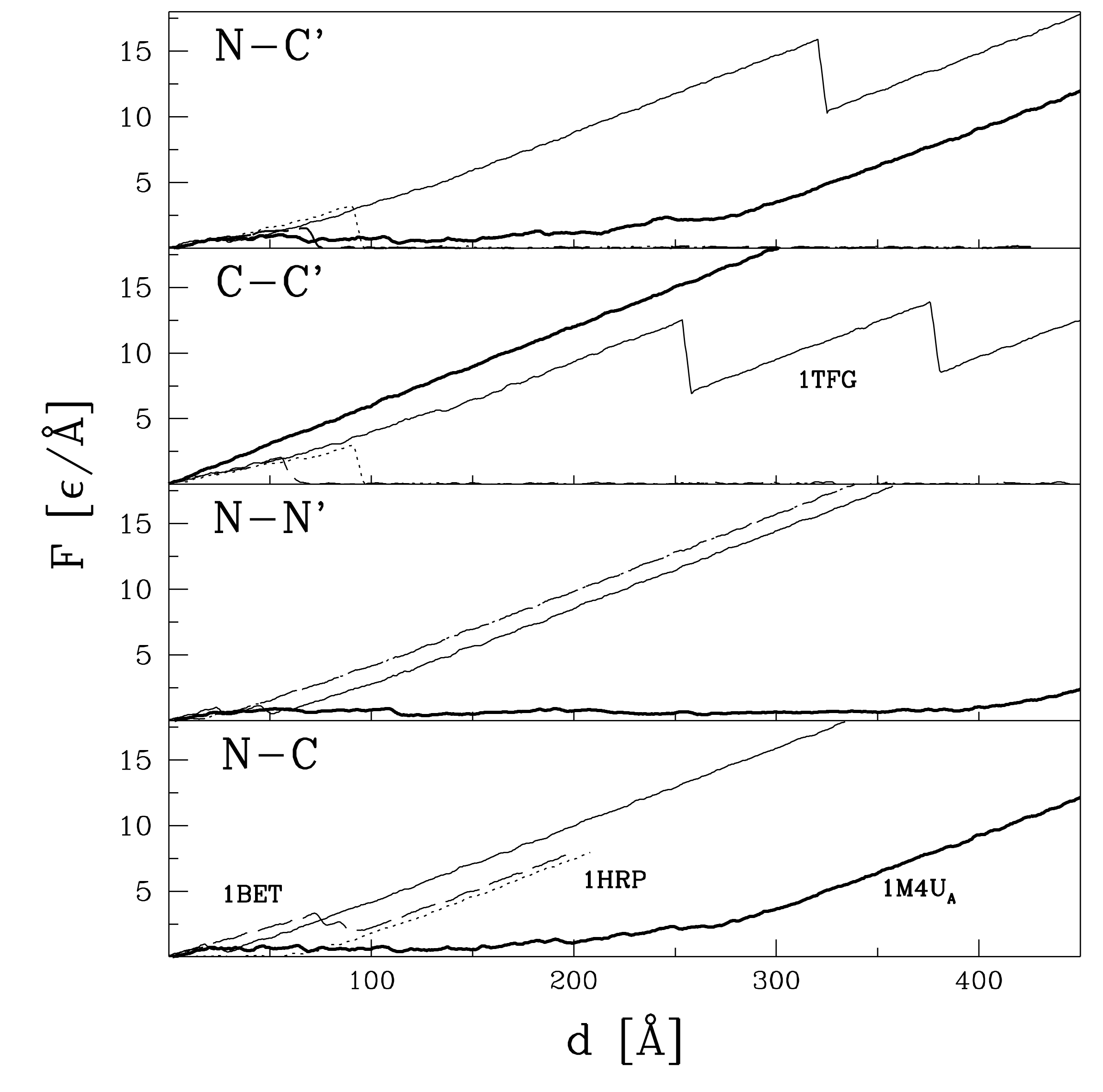}
\caption{{\small Similar to figures \ref{typ_1bmp} and \ref{typ_1fzv} but
for $\mathrm{1M4U_{A}}$
and the remaining proteins with cystine knots listed in Table I.}
} \label{inne_1m4u}
\end{figure}

The $F-d$ plots for noggin, shown in figure \ref{inne_1m4u},
are similar to those for the VEGF family (figure \ref{typ_1fzv}) in the sense
that no force peak develops in the C-C' scheme, but the force peaks appearing
within other schemes of pulling are minor.

We now consider proteins, 1BET and 1HRP, in which the
bridges between monomers are not cystinic. For the N-C pulling, $F_{max}$
is equal to 3.4 $\epsilon$/{\AA} in the case of 1BET, but there is no
force peak for 1HRP. For other kinds of pulling there is an eventual
separation of the monomers which is preceded by development of
a force peak resulting from overcoming the shear.
Even though the cystine ring in 1BET consists of more residues than in 1HRP,
the largest force peaks for both are similar in height: 3.4 $\epsilon$/{\AA}.
These force peaks are related to the shear
between the monomers and they do not involve formation of any slipknot.

The $F-d$ curves for the monomeric knottins are shown in figure
\ref{conotoxin}. Each such curve comes with a single peak which
involves dragging of a slipknot through the ring. The largest
mechanostability in this set is displayed by 1JU8 (a 4-kDa peptide
found in legumes). The corresponding $F_{max}$ is about 5.8 $\epsilon$/{\AA}.
This is a remarkably large force, considering that 1JU8 consists of only 37 residues
(the survey in ref. \cite{PLOS} has dealt with proteins of at least 40 residues).

\begin{figure*}[htb]
\includegraphics[width=0.9\textwidth]{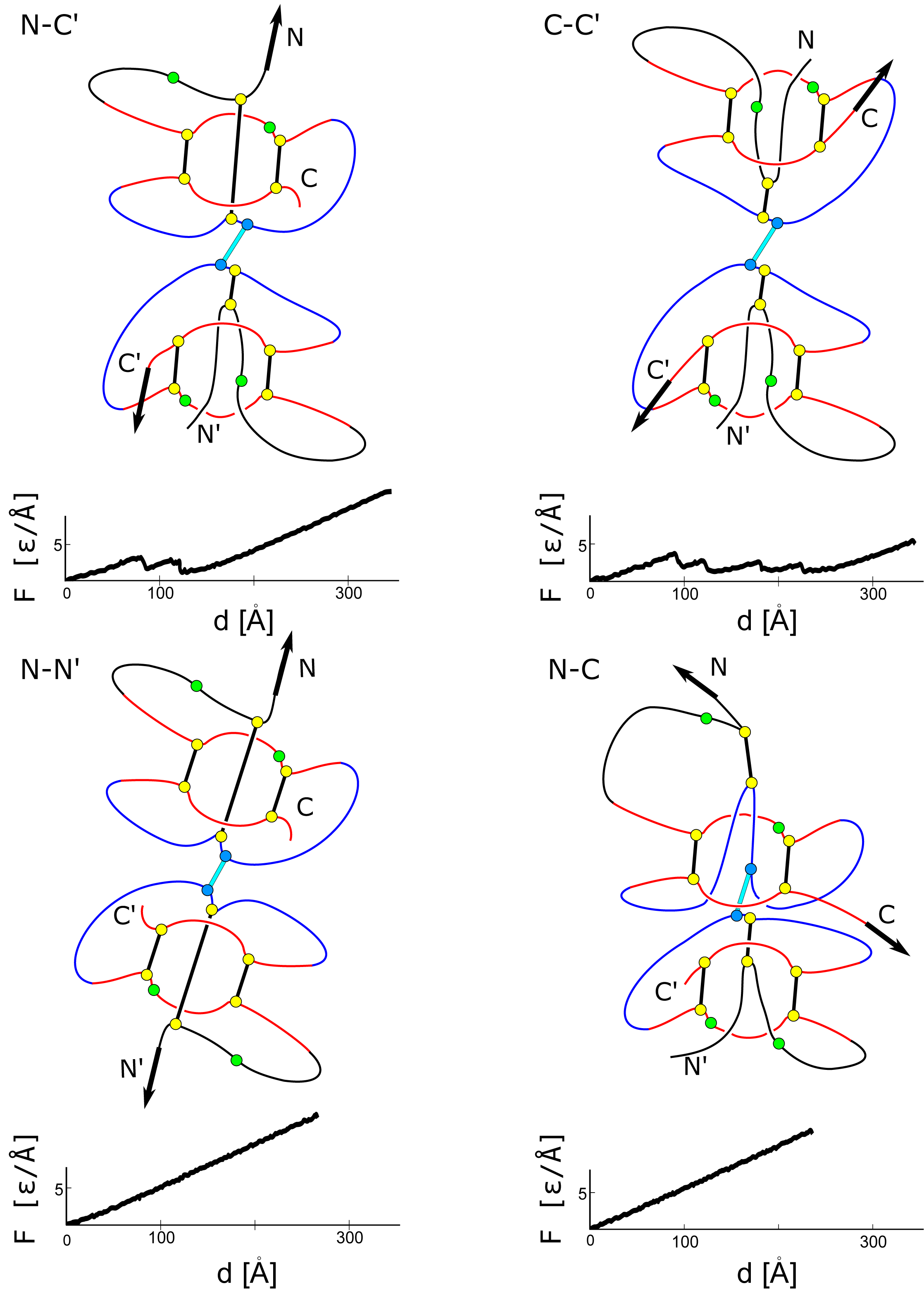}
\caption{{\small Mechanisms involved in stretching of a protein from the TGF family
for the four choices  of attachment points. Below each panel, there is a
corresponding $F-d$ plot obtained for 1BMP.}
}\label{1bmp_wspolny}
\end{figure*}

\begin{figure*}[htb]
\includegraphics[width=0.9\textwidth]{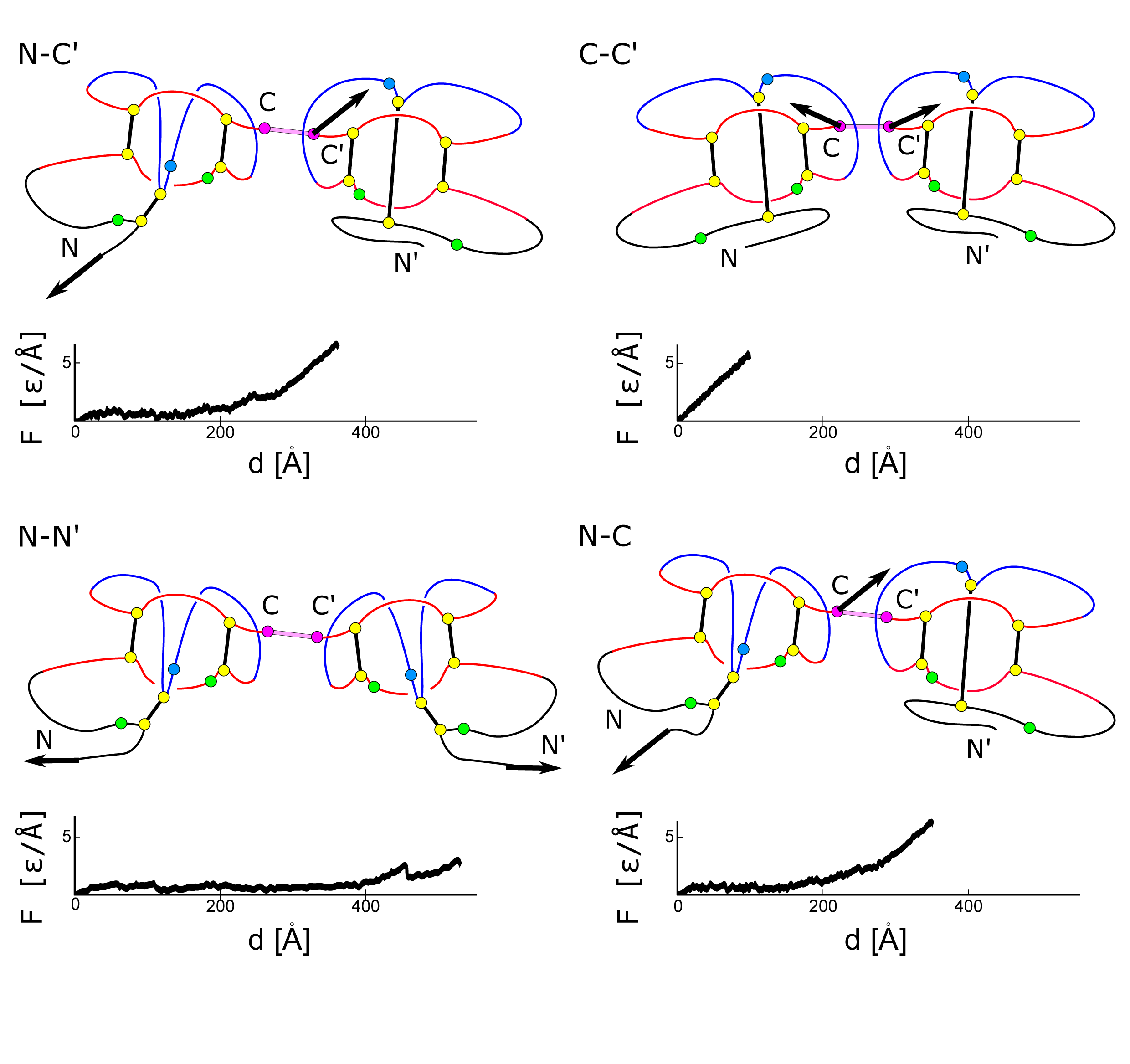}
\caption{{\small Mechanisms involved in stretching of noggin ($\mathrm{1M4U_{A}}$)
for the four choices of attachment points. Below each panel, there is a
corresponding $F-d$ plot.}
}\label{1m4u_wspolny}
\end{figure*}

\begin{figure*}[htb]
\includegraphics[width=0.9\textwidth]{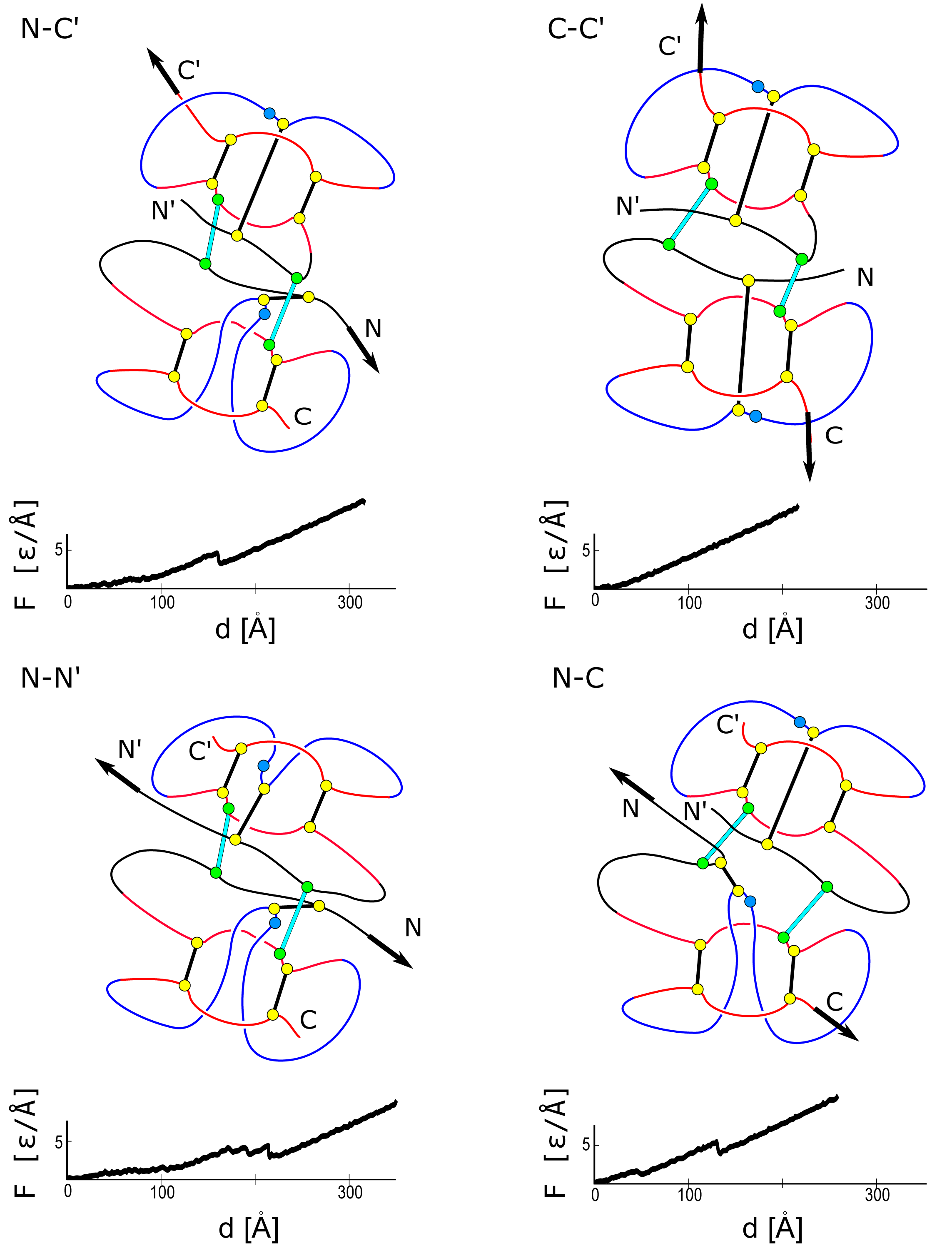}
\caption{{\small Similar to figures \ref{1bmp_wspolny} and \ref{1m4u_wspolny}
but for a protein from the VEGF family. The $F-d$ plots are for 1FZV.}
}\label{1fzv_wspolny}
\end{figure*}

The mechanisms involved in stretching are illustrated in figures \ref{1bmp_wspolny},
\ref{1m4u_wspolny}, and \ref{1fzv_wspolny} for 1BMP (representing TGF),
1M4U$_{\rm A}$, and 1FZV (representing VEGF) respectively.
The figures also show the corresponding $F-d$ plots.
We first consider stretching of 1BMP.  There are no force peaks for the N-N' and
N-C ways of pulling as the tension builds up, indefinitely, only in the covalent
bonds (along the backbone and in the disulfide bonds)
-- see also figure \ref{ineffective}.
In the case of the N-N' stretching, the system stiffens immediately on pulling
as the ring piercing cystines and the monomer connecting cystine align in series
(the lower left panel of figure \ref{1bmp_wspolny}) without rupturing
any native contacts and without formation of a slipknot -- only the covalent
bonds get stretched. In the case of N-C, the situation is more subtle.
The slipknot in the first monomer cystine ring does form but it does not
go all the way through to generate a force peak before initiating a process
of tension relaxation after overcoming a bottleneck.
The reason is that dragging of the slipknot is halted by the second monomer
which would have to go through the ring, but it is too big to squeeze in.
Pulling in the remaining cases, N-C' and C-C', generates articulated force peaks
because the pulling direction is not parallel to the cysteine that connects
the monomers. This results in non-simultaneous passages of two slipknots.
In the case of C-C' (see also figure \ref{snap},
the force peaks are split further because of
shearing in a $\beta$-sheet near the N-termini.
After the pivotal part of the knot loop squeezes
through the ring, it is held for an instant by the hydrogen
bonds between strands $A_1$ and $A_2$ shown in figure \ref{structure_razem} 
These contacts need to be broken to negotiate threading of
the knot loop through the ring.
The same scenario is observed in the $C-C'$ pulling.
However, it is repeated twice as both monomers act symmetrically.

$\mathrm{1M4U_{A}}$ is structurally similar to the TGF proteins but
the monomers are linked at the C and C' termini which results
in no force peaks for the C-C' way of pulling. For the remaining modes of
stretching, one or two (in the case of N-N') slipknots form but the resulting
peaks are barely observable (see figure \ref{1m4u_wspolny}).
This is because the cystine ring consist
of 10 residues instead of 8 and thus provides less hindrance to
the motion of a slipknot.

The mechanisms of stretching in the VEGF proteins are outlined in
figure \ref{1fzv_wspolny}. Notice that the dimer has a two-fold
symmetry with respect to the axis perpendicular to the plane containing
all four termini (thermal fluctuations make this symmetry approximate).
Thus pulling within
schemes N-N' and C-C' results in a motion which is symmetrical with
respect to the axis and in equal tensing of the two inter-monomer bridges.
In the case of N-N' this means nearly simultaneous generation of
two slipknots whereas in the case of C-C' no force peaks.
In the other two cases, there is no symmetry and a single slipknot
is generated leading to a noticeable force peak.

\subsection*{Comparison to single chain stretching}

\begin{figure}[htb]
\includegraphics[width=0.5\textwidth]{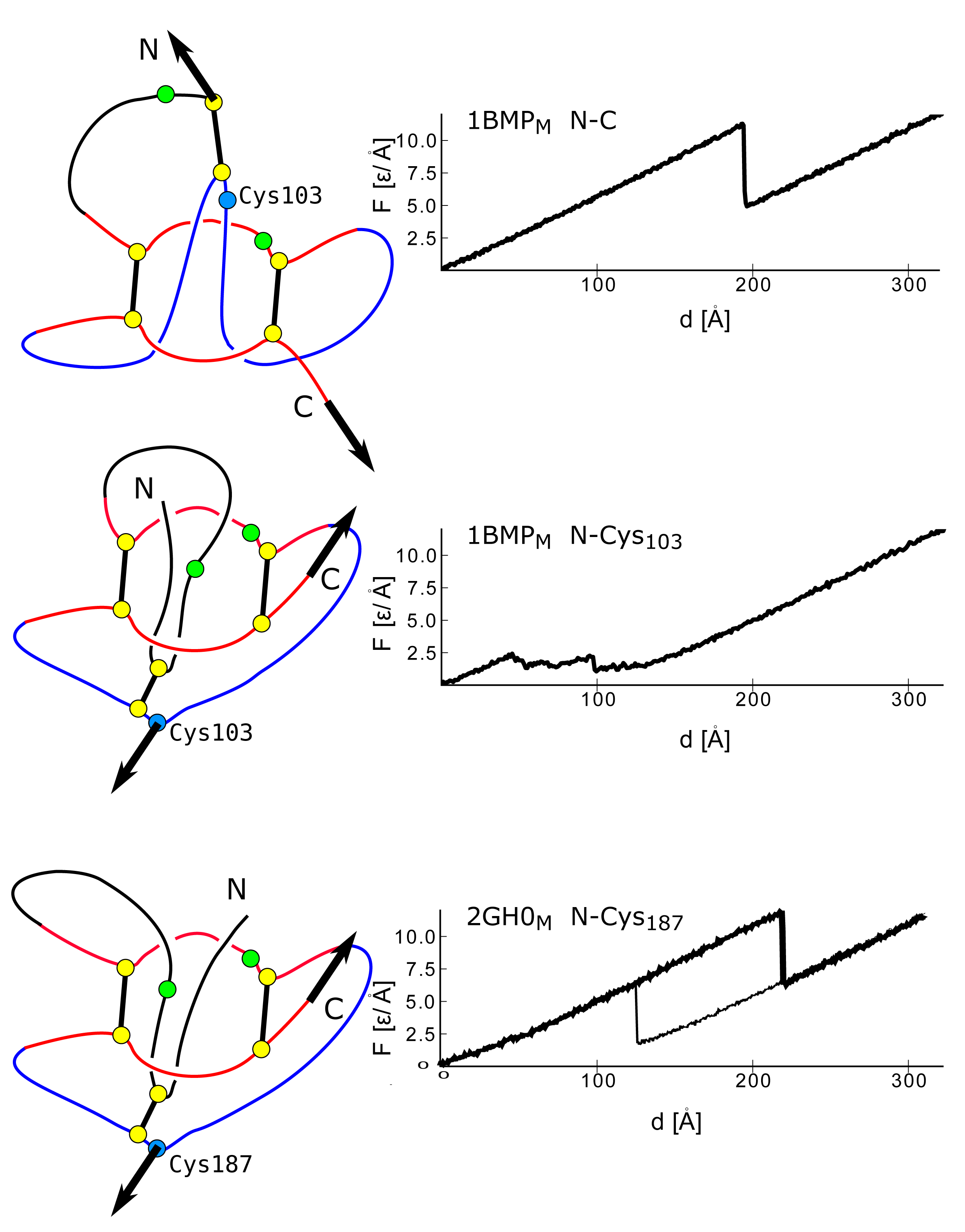}
\caption{{\small Top two rows of panels: Comparison of pulling of the monomeric 1BMP
at different points of attachment of the pulling force.
Bottom row of panels: stretching of the monomeric 2GH0. The thick force line is
for the N-Cys187 pulling and the thin force line is for a similar situation,
in which, however, the contacts between the knot-loop (strands $B_{1}$ and $B_{2}$
in figure \ref{structure_razem}) and the rest of the protein is removed.
These contacts affect the angle at which the ring piercing cystine is dragged
across the ring: they make the pulling at a small angle between the plane of
the ring and the plane of the slipknot. In the absence of these contacts,
the approach is more vertical which results in the observed reduction of the force.
}}\label{1bmp_old_vs_new_vs_2gh0}
\end{figure}

\begin{figure}[htb]
\includegraphics[width=0.5\textwidth]{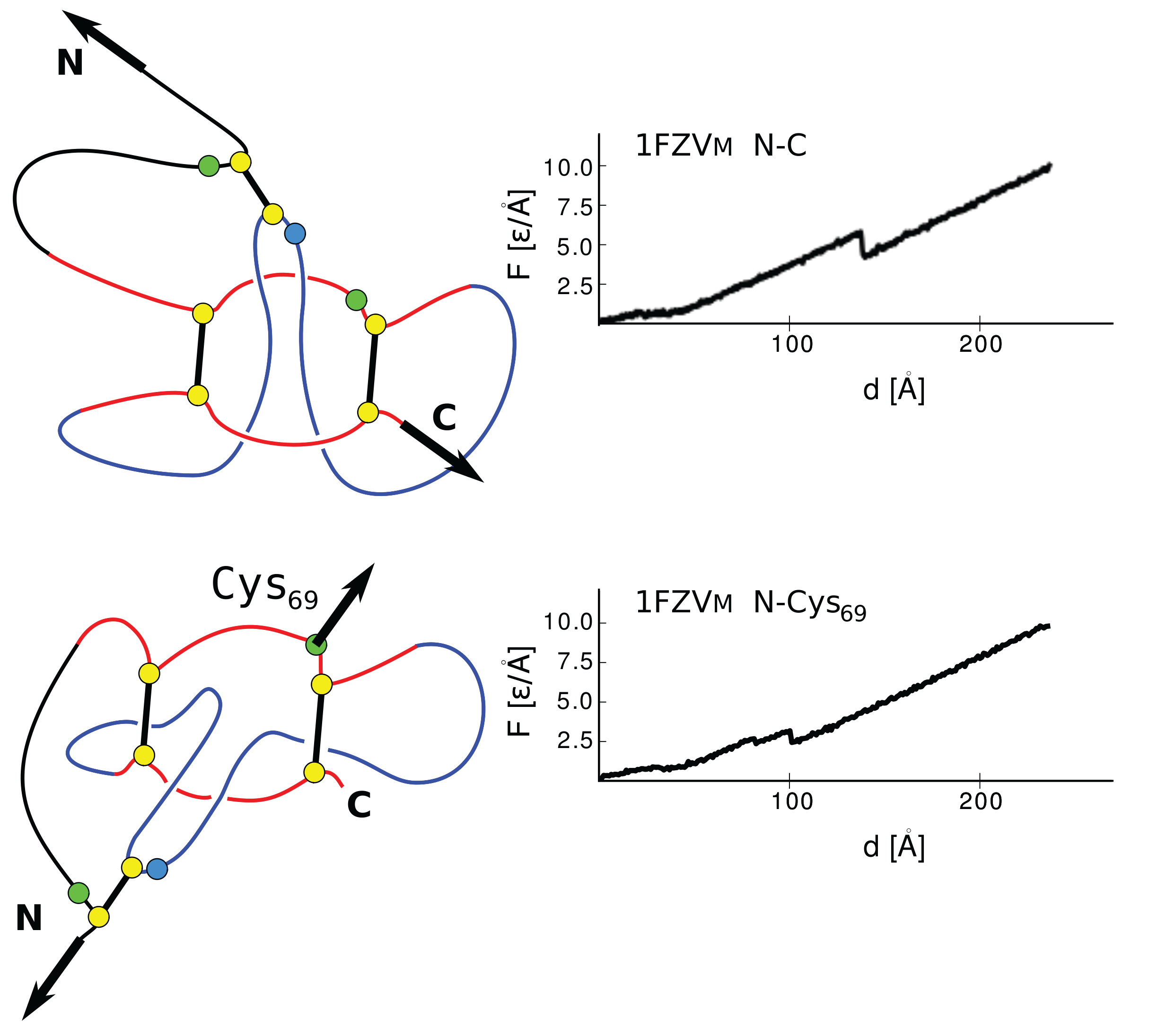}
\caption{{\small Stretching of the monomer of 1FZV at different points of
force attachment.
The top panel shows the conventional stretching of a monomer by N and C termini.
This kind of manipulation is inaccessible
when the protein is in its active, dimeric state.
Stretching in the dimeric N-C case, however, results in effectively
shifting the location of the pulling force from the C-proximal cysteine
on the ring to another cystine, Cys69, on this ring.
The deformation of the ring is different and, in addition, the slip-loop
crosses the ring more vertically. Both of these factors result in an
overall reduction in the force in the dimeric N-C stretching.
}}\label{1fzv_old_vs_new}
\end{figure}

Table I shows the values of $F_{max}$ for the dimeric situations but also
compares them to those obtained assuming that a monomer is not connected
to any partner. For some proteins (2GH0, 1M4U$_{\rm L}$, 1TFG, 1WQ9, 1VPF)
the monomeric values of $F_{max}$ are smaller than the largest $F_{max}$
for the peak-bearing dimeric cases. For some (2GYZ, 1FZV) -- almost the same.
In the remaining cases we observe lowering of $F_{max}$ -- even by
a factor of 2 as for 1BMP.
The short explanation for the observed differences
is that even though monomeric and dimeric peak forces are due to cystine
slipknots, dragging of the knot-loop usually takes place in different directions
(e.g. opposite) or different angles with respect to the cystine ring.
These circumstances can be explained with the help of
figures \ref{1bmp_old_vs_new_vs_2gh0} and \ref{1fzv_old_vs_new} for
1BMP and 1FZV respectively. The top panels refer to the monomeric
pulling in the N-C way which produces dragging of the slipknot towards
the N terminus which is sequentially close to one partner of the
cystine that does the dragging.

In the dimeric 1BMP, the force peaks arise in the N-C' and C-C' stretchings
which results in dragging of the slipknot by Cys103 in the opposite
direction and in pulling of the N-terminus into the ring. The effect on
the $F_{max}$ is a four-fold reduction in the value when this kind
of manipulation is induced in the monomeric 1BMP.
The reduction is also related to a more vertical dragging of the slipknot
through the ring which results in a smaller deformation of the ring.
The double peak is due to letting the N-terminus that is followed by
sliding of the knot-loop through.
The situation is quite
similar for other TGF proteins. A bigger difference occurs for 2GH0, where
Cys187 acts as the pulling agent. This difference is that the N-terminus
does not enter the ring on the rim side. Instead, the structure of the
protein is such that the slipknot is rotated so that the N-terminus
enters near the center, requiring more space to pass without
separating the passage into the distinct knot-loop and N-terminus events.
In this case, $F_{max}$ gets doubled.
The physics of the slipknot formation in 1TFG is quite special as it
involves dragging of a ring through a ring which we dubbed the
mechanism of the cystine plug. We deal with it
with in a separate publication \cite{plug}.

For the N-C' pulling of the dimeric 1FZV, the relevant forces are effectively
attached to N and Cys69 on the cystine ring.
(Cys69 transports most of the tension between the monomers.)
The resulting slipknot
still protrudes towards the N-terminus, as in the N-C pulling of the monomer,
but it makes a different angle with respect to the ring (the bottom panels
of \ref{1fzv_old_vs_new}) which results in a reduction of $F_{max}$.


We now come back to figure \ref{1fzv_wspolny} and observe that when
the dimeric 1FZV is stretched in the symmetric N-N' way, there are,
surprisingly, three consecutive force peaks. Judging by the symmetry,
we would expect to find an even number of peaks. We could explain
this puzzle by a selective switching of certain contacts off
and investigating the effect of such an  action.
Each monomer of this protein encompases
two characteristic $\beta$ sheets, each containing two
strands, as shown in Fig.~\ref{structure_razem}  
In 1FZV, the longer sheet (strands $C_1$ and $C_2$) spans the whole length of
the protein comprising 17 amino acid on each of the strands.
When the protein is dimerised, the sheet is facing the outer side
of the complex. The shorter of the $\beta$ sheets ($B_1$, $B_2$)
runs parallel to the long one and contains 8-10 amino-acids.
In the dimeric state, it faces the inter-monomer interface
but it does not form a larger sheet with its symmetric counterpart.
The two sheets are bridged together by a cystine ring.
We have found, that
shearing of the bonds within the shorter stretch of
$\beta$ sheet is responsible for the emergence of the
first and smaller peak. This process arises in both
monomers simultaneously.
Indeed, the first of the peaks disappears upon
removal of contacts between the two shorter $\beta$
sheets in each monomer ($B$ and $B'$).
Clearly, these contacts are not only responsible
for the first peak, but they also contribute to the
remaining two peaks, as their removal reduces the
height of the remaining two peaks.
To summarise, in the monomeric 1FZV, $F_{max}$ is almost entirely
due to steric hindrance but in the dimeric 1FZV there is also
a contribution from attractive native contacts.

\subsection*{Monomeric cystine knot proteins}

\begin{figure}[htb]
\includegraphics[width=0.5\textwidth]{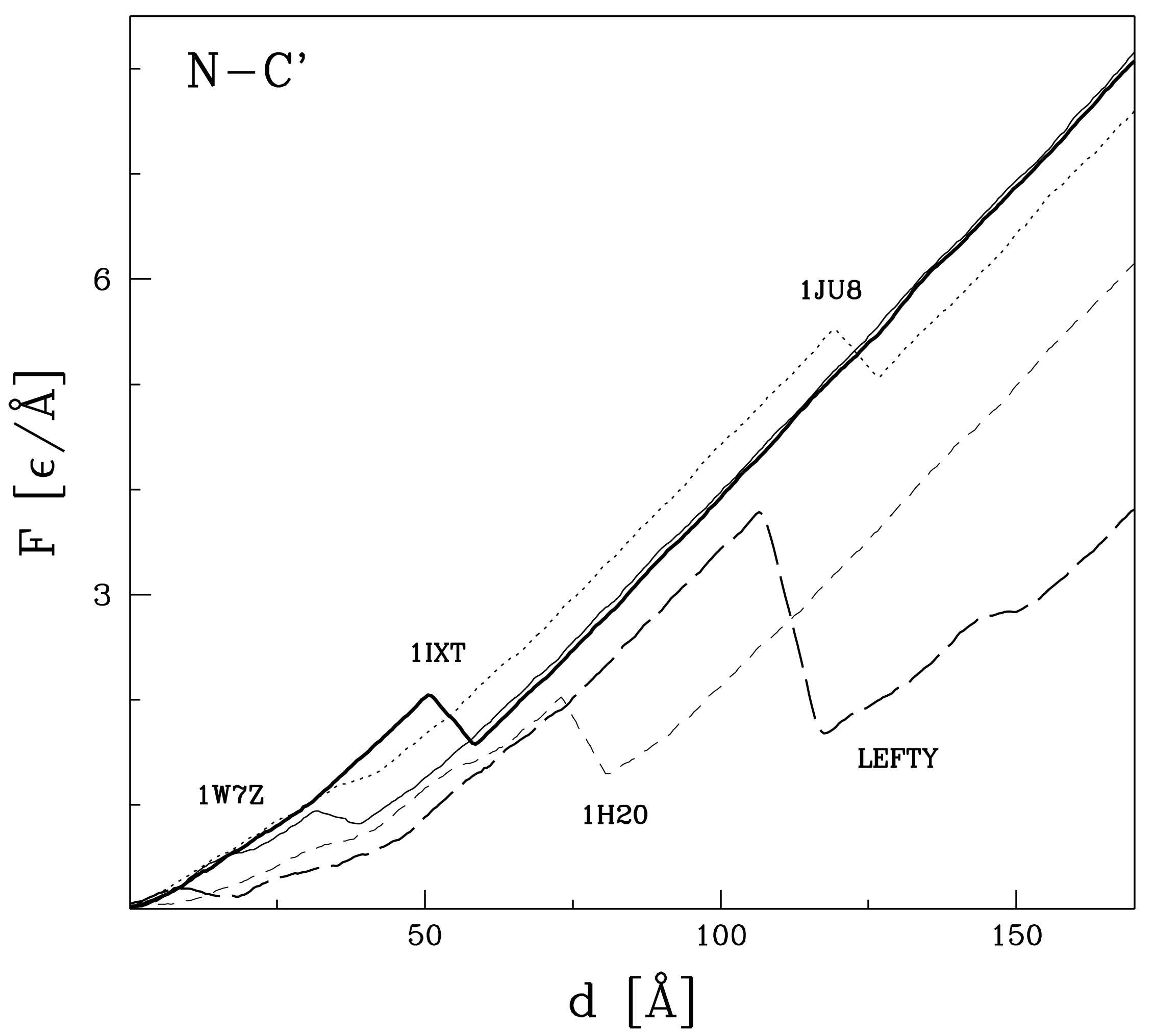}
\caption{{\small The $F-d$ curves for the monomeric proteins with a cystine knot that are listed
in Table I. The stretching is accomplished by the termini.}
} \label{conotoxin}
\end{figure}

We now consider the cystine knot proteins that are monomeric.
One example is knottins. Figure \ref{conotoxin} shows the $F-d$ curves
for four examples of such proteins.
The first example is a toxin extracted from from a sea snail
{\it Conus gloriamaris}\cite{conotoxin}. This protein has the structure code
1IXT and it comprises only 27 amino-acids. Structurally, it is
related to the family of cyclic disulfide-rich peptides.
Despite its short length, our simulations of stretching suggest existence
of an articulated force peak with $F_{max}$ of order 2.2 $\epsilon$/{\AA}
which just exceeds $F_{max}$ of titin when calculated within the same model \cite{PLOS}.
The second example is the tripsin inhibitor protein II (code 1W7Z)
derived from the poisonous squirting cucumber {\it Ecballium elaterium}.
which shares the cystine-knot topology. In contrast,
it yields only a minor force peak with $F_{max}$ of 1.2 $\epsilon$/{\AA}.
The smaller force is due to the larger size of the cystine ring -- it
consists of 11 residues instead of 8 as in 1IXT.
The 39-residue potato carboxypeptidase with the code 1H20 is found to be as
strong as 1IXT. The biggest
$F_{max}$ in the set considered is predicted to arise in the 31-residue long leginsulin
(code 1JU8). The corresponding  $F_{max}$ of about 5.8 $\epsilon$/{\AA}  is
as big as for the most robust dimeric cystine knot proteins listed in Table I.

Another example of monomeric cystein knot systems is
a group of proteins from the nodal signaling pathway which
governs cell differentiation in embryonal development \cite{schier} such as lefty,
for which an existence of monomers was suggested.
No structure of lefty protein has been solved to date.
We have thus resorted to a homology model, that has been calculated
using ModPipe software\cite{homologymodel} using a standard set of parameters.
It should be noted that the disulfide bridge between cysteins 251 and 253
in the derived structure was above a distance treshold and had to be set manually.
We predict that $F_{max}$ for lefty (see figure \ref{conotoxin}) should be
nearly twice as big as for conotoxin: $F_{max}$=4.1$\epsilon$/{\AA}.
In all examples considered in this section, the force peaks arise due to
formation of the cystine slipknot conformation.
The hypothesis that growth factors form dimers due to the
lack of the hydrophobic core appears not to be valid in the case of the monomeric lefty, 
which is the member of the same family. However, no structure of this protein
has been solved so the judgement should be suspended.


\subsection*{Non-mechanical stability}

We now come back to the dimeric cystine knot proteins.
One may compare the largest value of $F_{max}$ obtained in one of the four ways
of stretching the dimers to $F_{max}$ derived for stretching of an extracted
monomer. By looking at the values listed in Table I, we realize that the difference
between the two forces, $\Delta F_{max}$, can be either positive or negative.
In contrast, as evidenced by the last column, the thermodynamic stability
resulting from joining two monomers into a dimer is always enhanced.
The change in thermal stability, $\Delta T_f$, is defined as the difference in
$T_f$ between the dimer and two separate monomers.
The degree of this enhancement is not related to $\Delta F_{max}$.
For instance, the largest $\Delta T_f$ of 0.04 $\epsilon/k_B$ (about 35 K) is
for 1QTY and 1VPF whereas $\Delta F_{max}$ for these two proteins is 3.3 $\epsilon$/{\AA}
and merely 0.6 $\epsilon$/{\AA} respectively. The values of  $\Delta T_f$ were inferred based on five long trajectories.
Our results are related to
the outcome of experiments on the VEGF proteins carried out by Muller
et al. \cite{Muller}. The authors have mutated particular cysteines within the
cystine knot and found out that this action results in worse folding and somewhat
reduced thermal stability. Here, on the other hand, we assess the effect of
dimerization on thermal stability of the GFCK proteins and conclude that the
inter-monomer disulfide bridges have similar effect, i.e. they increase
the thermal stability. Furthermore, a two-bridge connectivity (as in the TGF proteins)
tends to provides  a bigger thermal stability than a one-bridge connectivity
(as in the VEGF proteins). Another way to see the role of the cystins is by reducing
all of the disulfides. This has been  accomplished for artemin (2GH0) \cite{Bruinzeel}.
The resulting lowering of the folding temperature was of order 40 K.
Yet another study investigates the role of inter-monomer disulfide
bridges~\cite{denaturation1,denaturation2} in PDGF. It has been found on mutating
the cysteines in the intermeric disulfide bridges into serines
does not prevent the dimer from forming. However, its resistance
to chemical denaturation and changes in pH is reduced dramatically.

The TGF-$\beta$ proteins are known to display very slow chemical
denaturation \cite{bioslip1,bioslip2}. It may last for hours
and refolding is even slower. The corresponding rates
increase linearly with temperature. Interestingly, the authors
of ref. \cite{bioslip1} have proposed an unfolding mechanism
which is essentially similar to the cystine slipknot clamp
but defined for chemical/thermal unfolding. In their mechanism,
the slip-loop is slowly dragged out of the cystine ring as a result
of entropic effects. In our model of mechanical stretching,
the time scales are several orders of
magnitude too short to observe such fluctuational processes.
On the other hand, the slow nature of the refolding processes
seen experimentally explains the irreversibility of unfolding
observed in our studies.

\subsection*{Concluding remarks}

In this paper, we have shown that the primary mechanical clamp
associated with the force peaks in proteins with the cystine knots
is formation of the cystine slipknot, independent of whether the proteins
are  monomeric or dimeric.  
In our calculations, we include no attractive non-native
contacts in the model. Whereas they might play a role in folding,
their effect can only be minor in the context of the cystine
slipknot mechanism which is dominated by steric interactions.
We have elucidated the workings
of the cystine slipknot mechanical clamps in dimeric systems
and demonstrated emergence of interesting topological transformations.
We have shown that dimeric systems with the cystine knot should be
giants of mechanostability like the corresponding monomeric systems,
but the picture is more subtle since generating large force
peaks requires stretching only in certain directions.
Furthermore, the action of this kind of clamp in dimers
may be different from that in monomers. For instance,
dragging of a single slip-loop may take place in different direction
than in the corresponding monomer and the resulting $F_{max}$
may be either smaller or larger than in the monomer.

It is probably unlikely that forces of this magnitude affect
proteins with the cystine knots under biological conditions.
However, our studies may motivate research in bio-inspired materials
containing such proteins as building blocks. These materials would behave
similar to the spider dragline as they could absorb and
dissipate large ammounts of energy. Simple polymers, even if
very stiff, do not develop force peaks on the $F-d$ curves
whereas the energy which is absorbed and then dissipated depends on the
area under the peaks. Crossing a peak locks the system (practically
irreversibly) in a stretched conformation and generates an
element akin to a sacrificial bond.
In this way, fibers or networks made of such proteins should be able
to withstand forces larger than those associated  with systems made of
simple polymers.

Exploration of such bio-inspired materials should be preceded
by an experimental verification of our findings on mechanostability.



\section*{Acknowledgments}
This work has been inspired by a discussion with Alex Wlodawer.

\bibliography{template}

\section*{Figure Legends}

\begin{figure}[htb]
\includegraphics[width=0.5\textwidth]{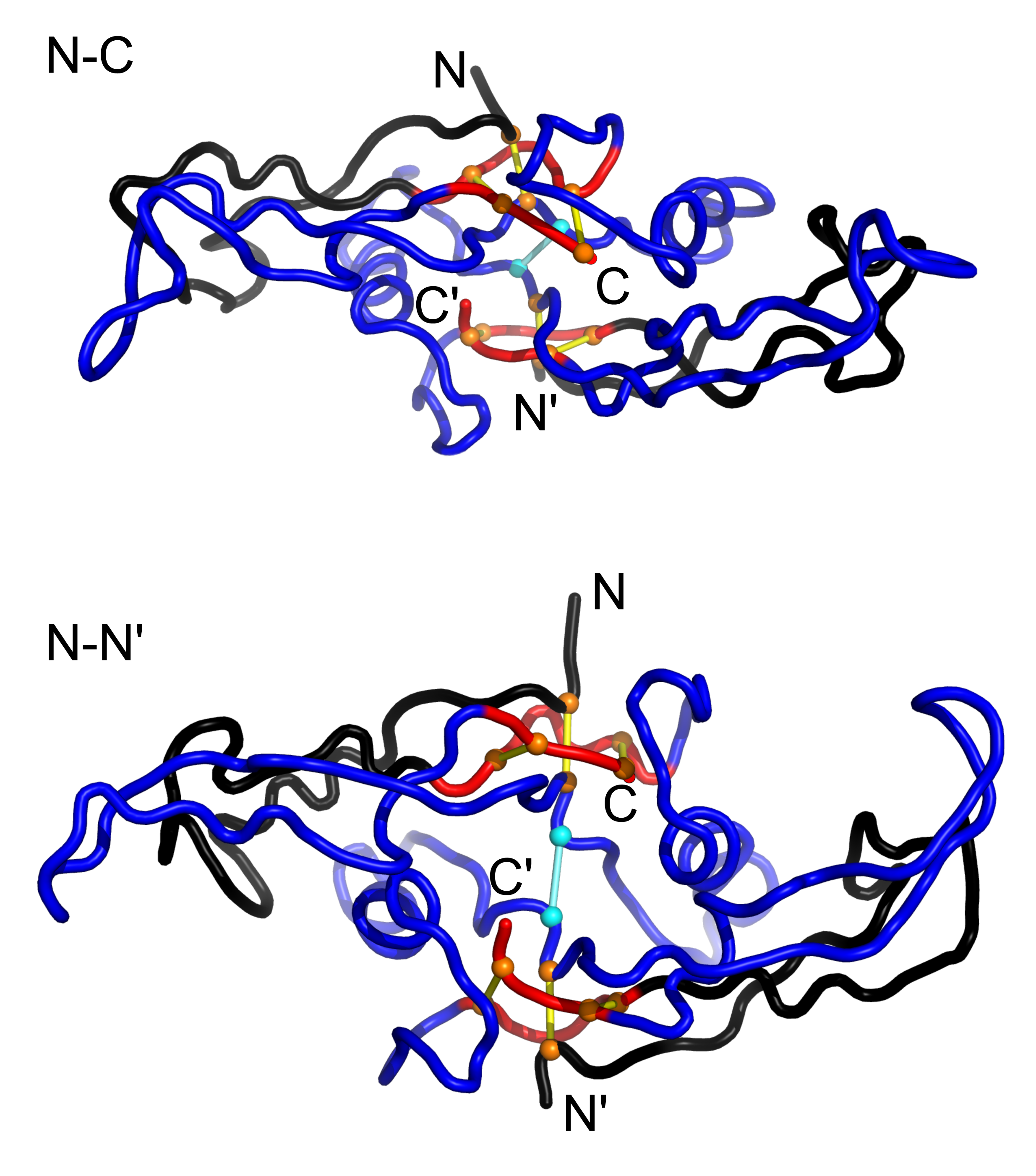}
\caption{{\small Examples of stretched conformations of 1BMP for the
NC (top panel) and NN' (bottom panel) pulling at $d$=250 {\AA}.
peaks arise through manipulations of these kinds.
The color coding is similar to that used in Figure \ref{dimer_types}.
In the top panel, the N
terminus pulls on the blue loop of the first monomer. The resulting
force is transferred to the second monomer through the inter-molecular
cystine bridge that is shown in cyan. The second monomer is too big
to cross the cystine ring and, therefore, the tension grows indefinitely,
exceeding values needed to break covalent bonds. In the bottom panel,
stretching results in an immediate allignment of the three cystine
bridges: within the cystine knots (in yellow) and the intermolecular one,
and in an indefinite growth of the tension.
}}
\label{ineffective}
\end{figure}

\begin{figure}[htb]
\includegraphics[width=0.5\textwidth]{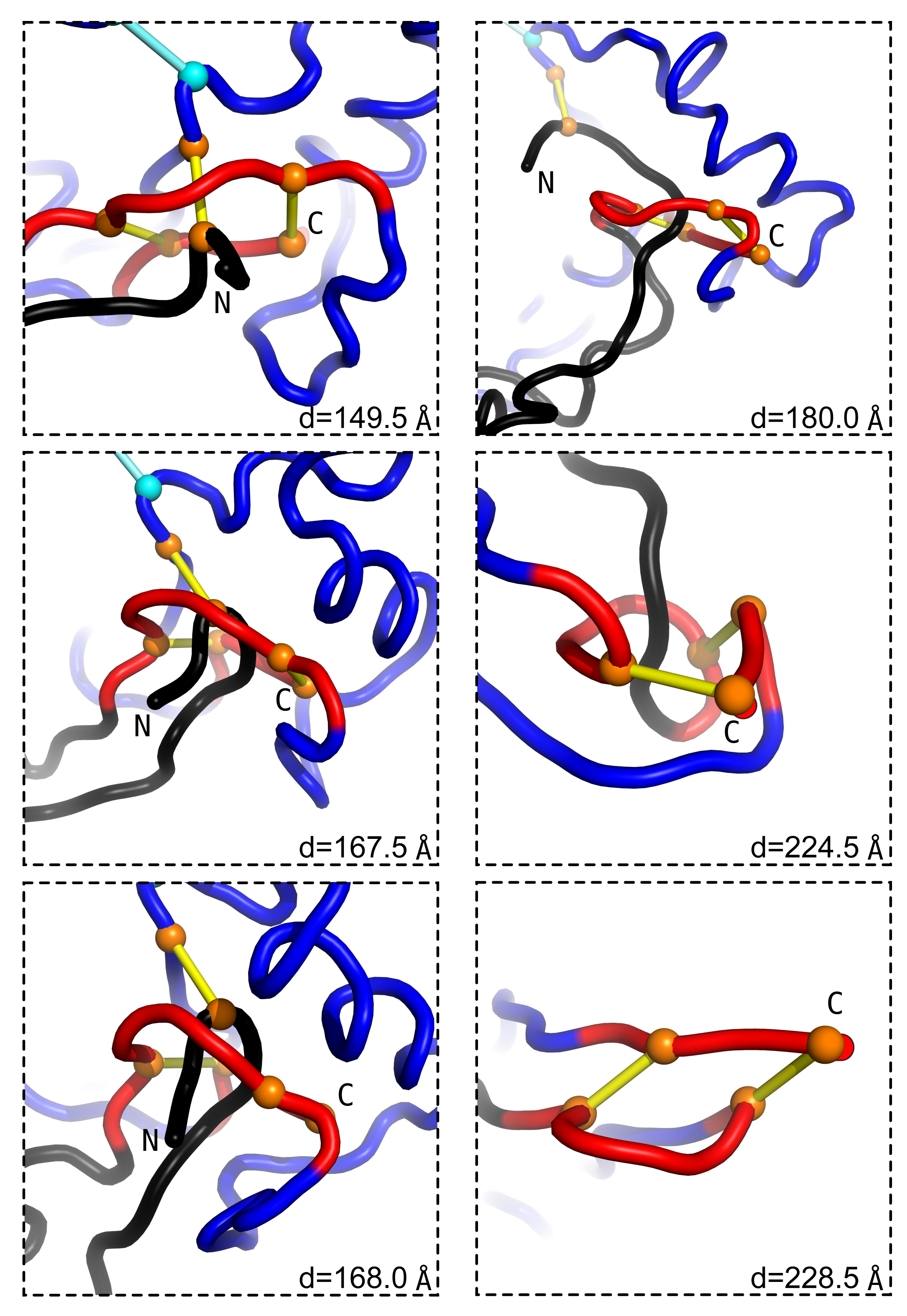}
\caption{{\small Subsequent snapshots of the model 1BMP during the
C-C' stretching. The corresponding values of $d$ are indicated.
The figure shows only the region in which the cystine slipknot forms.
The first of the frames shows the knot near it's native state. In the
middle left panel the knot loop (shown in black) approaches the
inside of the ring. In the next snapshot, it squeezes halfway through the ring.
This is the stage corresponding to the highest tension reached during
unfolding. In the next frame (top right), the loop has already slipped
past the ring. At this point, the system is unable to return to the
native state rapidly if the pulling spring is removed.
Two subsequent frames show further extension of the protein.
In the bottom right panel, the whole length of the slip-loop has crossed the
ring: the slipknot is released.
The second force peak is due to the formation
of the cystine slipknot in the second monomer (not shown).
}}
\label{snap}
\end{figure}

\clearpage
\section*{Tables}
\clearpage
\begin{table}
\caption{{\small Values of $F_{max}$ of the proteins studied here in units of
$\epsilon$/{\AA} and for different pulling schemes.
$F_M$ denotes $F_{max}$ in the monomeric case -- when only
one chain of the dimer is considered in the N-C mode.
The four penultimate columns are for the the dimeric situation. The subscripts
of $F$ indicate the mode of pulling.
$S_r$ denotes the number of amino acids in the cystine ring.
The last column shows a difference between melting temperature of a dimer and
two separated monomers, $\Delta T_f$ in units of $\epsilon/k_B$}
}
\begin{tabular*}{\hsize}{@{\extracolsep{\fill}}lllp{1cm}lllll}
\label{tabela1}
PDBid & Family  & $\;\;S_r\;\;$ & $\;\;F_{M}$ & $F_{N-C'}$ & $F_{N-N'}$& $F_{C-C'}$ & $F_{N-C}$ & $\Delta T_f$    \cr
\hline
{\bf dimeric}       &        &            &         &       &       &       &                   &               \cr
1BMP                & TGF    &$\;\;$ 8    & 10.3    &  3.4  & $-$   & 4.0   &  $-$              &     0.02      \cr
1LXI                & TGF    & $\;\;$ 8   &  7.3    &  4.0  & $-$   & 3.5   &  $-$              &     0.02      \cr
2BHK                & TGF    & $\;\;$ 8   &  7.3    &  3.5  & $-$   & 3.7   &  $-$              &     0.01      \cr
2GH0                & TGF    & $\;\;$ 8   &  5.9    &  8.6  & $-$   &12.0   &  $-$              &     0.03      \cr
2GYZ                & TGF    & $\;\;$ 8   &  5.4    &  4.6  & $-$   & 5.5   &  $-$              &     0.03      \cr
1REW                & TGF    & $\;\;$ 8   &  5.3    &  4.5  & $-$   & 3.2   &  $-$              &     0.02      \cr
1M4U$_{\rm L}$      & TGF    & $\;\;$ 8   &  5.3    &  8.0  & $-$   & 8.0   &  $-$              &     0.01      \cr
1TFG                & TGF    & $\;\;$ 8   &  5.5    &  14.2 & 1.1   &14.0   &  1.1              &     0.02       \cr 
1QTY                & VEGF   & $\;\;$ 8   &  8.9    &  5.6  & 4.6   & $-$   &  1.9              &     0.04      \cr
1CZ8                & VEGF   & $\;\;$ 8   &  6.4    &  5.7  & 4.9   & $-$   &  5.8              &     0.04      \cr
1FLT                & VEGF   & $\;\;$ 8   &  5.5    &  5.2  & 4.4   & $-$   &  4.7              &     0.03      \cr
1WQ9                & VEGF   & $\;\;$ 8   &  5.5    &  4.5  & 4.7   & $-$   &  6.1              &     0.03      \cr
1FZV                & VEGF   & $\;\;$ 8   &  5.4    &  4.5  & 4.2   & $-$   &  5.5              &     0.03      \cr
1VPF                & VEGF   & $\;\;$ 8   &  5.3    &  5.9  & 5.0   & $-$   &  5.8              &     0.04      \cr
$\mathrm{1M4U_{A}}$ & noggin & $\;$ 10    &  2.8    &  2.5  & 2.6   & $-$   &  1.5              &     0.02      \cr
1BET                & NGF    & $\;\;$ 14   &  2.1    &  1.6  & 2.4   & 2.1   &  3.4              &     0.02      \cr 
1HRP                & HCG    & $\;\;$ 8   &  $-$    &  3.4  & 2.0   & 3.0   &  $-$              &     0.02      \cr 
\hline
{\bf monomeric}     &        &            &         &       &       &       &                   &           \cr
lefty             & TGF       & $\;\;$ 8  &  4.1    &       &       &       &                   &           \cr
1IXT                & knottin & $\;\;$ 8  &  2.2    &       &       &       &                   &           \cr
1W7Z                & knottin   & $\;\;$ 11 &  1.2    &       &       &       &                 &           \cr
1H20                & knottin       & $\;\;$ 9  &  2.2    &       &       &       &             &           \cr
1JU8                & knottin & $\;\;$ 8  &  5.8    &       &       &       &                   &           \cr
\hline
\end{tabular*}
\end{table}

\clearpage
\section*{Supplementary Information}

The first movie (movie\_1.avi) shows
unfolding of the 1BMP dimer in the C-C' mode.
One monomer is colored in blue and the other in red.
The cystine bonds are indicated as yellow sticks.
The arrows indicate amino acids through which the pulling process is implemented.
The whole duration of the movie corresponds to 100 1000 $\tau$, where $\tau$
is of order 1 ns.

\vspace*{0.5cm}

The second movie (movie\_2.avi) is a close-up that is focused on the workings
of the cystine slipknot clamp in action. The coloring of the
backbone segments is the same as used in the main text. Halfway through,
an attempted but unsuccessful passage of the slip-loop can be seen.
The attempt becomes successful only
after the tension crosses the threshold value.
The movie ends at about half of the total duration
of the first movie, when all slip-loop has crossed the ring and no
further structural rearrangement is possible.

\end{document}